\documentclass[12pt]{article}
\usepackage{newcent}

\usepackage[T1]{fontenc}
\usepackage[utf8]{inputenc}
\usepackage[a4paper]{geometry}
\usepackage{color}
\usepackage{amsmath}
\usepackage{amssymb}
\usepackage{setspace}
\usepackage{esint}
\usepackage[unicode=true,pdfusetitle,
 bookmarks=false,
 breaklinks=false,pdfborder={0 0 1},backref=false,colorlinks=true,pdfpagemode=FullScreen]
 {hyperref}

\makeatletter
\newcommand{\lyxaddress}[1]{
\par {\raggedright #1
\vspace{1.4em}
\noindent\par}
}

\makeatother

\begin{document}

\title{Construction of dynamical invariants for the time-dependent harmonic
oscillator with a time-dependent driven force}

\author{M. C. Bertin%
\thanks{mbertin@ufba.br%
}, B. M. Pimentel%
\thanks{pimentel@ift.unesp.br%
}, J. A. Ramirez%
\thanks{johnarb@ufba.br.br%
}}

\maketitle

\lyxaddress{\begin{center}
$^{*}{}^{\ddagger}$Instituto de Física, Universidade Federal da
Bahia,\\
 Campus Universitário de Ondina, CEP 40210-340, Salvador, BA, Brazil.\\
$^{\dagger}$Instituto de Física Teórica, UNESP - São Paulo State
University.\\
 P.O.B. 70532-2, 01156-970, São Paulo, SP, Brazil.
\par\end{center}}
\begin{abstract}
\thispagestyle{empty}We construct the linear and quadratic polynomial
dynamical invariants for the classical and quantum time-dependent
harmonic oscillator driven by a time-dependent force. To obtain them,
we use exclusively the associated equations of motion for the system.
We also find an algebraic relationship between the linear and quadratic
invariants at the classical and quantum level. 
\end{abstract}

\section{Introduction\label{sec:Introduction}}

The driven time-dependent harmonic oscillator (driven TDHO) is one
of the most useful models in modern and classical physics. As a simple
model, it has several theoretical applications, \emph{e.g.} the study
of time evolution of quantum systems and its correlation with classical
mechanics \cite{Schuch}. Also, it can be used to study the optimization
of processes at the microscopic scale \cite{Muga}. Moreover, it has
experimental applications in areas such as molecular physics \cite{Muga1},
ion traps \cite{Manko,Glauber}, quantum computation \cite{Schleich,QCDI},
and accelerator physics \cite{Takayama}. 

Time dependent systems are not stationary, so the hamiltonian function
is not a conserved quantity, thus conventional approaches to deal
with time-independent systems cannot be used. Despite its explicit
time-dependence, the dynamical invariant approach allows an appropriate
analysis of the time-evolution of these systems in several situations,
offering the possibility to find solutions of the classical and quantum
dynamical equations. This analysis can be made in the cases in which
the study of the conserved quantities associated to the symmetries
of these systems can be implemented. In other words, in both classical
and quantum cases, complete integration can be performed by finding
its associated algebraic structure. That structure can be constructed
with the operators that belong to conserved quantities which can be
related to symmetries. The conserved quantities can be derived directly
from the differential equations, as shown in \cite{MBJ}, or by using
variational approaches \cite{Lutzky,Hojman}.

The principal characteristic of the algebrae associated with quadratic
systems, like the ones managed in this work, is the existence of linear
and quadratic operators associated to the dynamical invariants of
the system. In general, there are two linear invariants for each degree
of freedom. Using these operators, it can be generated a simple spectrum
which, despite not being associated with the energy, allows us to
find the solutions of the equations of motion \cite{Lewis}. 

Generally speaking, depending of the nature of the system, for each
case, classical or quantum, we can use several methods to construct
or find those dynamical invariants. Besides of the method proposed
by the authors in \cite{MBJ}, alternative procedures to obtain the
dynamical invariants can be used depending on the system that is dealt
with. For the quantum case we have Manko's \emph{et al.} \cite{Manko,Manko1}
and the Lewis' procedures \cite{Lewis}, and for the classical case
we can find Lutzky's and Hojman's ones \cite{Lutzky,Hojman}.

In this work, we provide a continuation of the work \cite{MBJ}, showing
how to construct the dynamical invariants for the time dependent harmonic
oscillator driven by a time dependent force, and deduce the relations
between the linear and quadratic dynamical invariants for the classical
and quantum problem. In sec. \ref{sec:classical} we perform the derivation
of linear and quadratic invariants from the classical equation of
motion, and show that the quadratic one can be put in the form of
product of two linear operators. The same is done in sec. \ref{sec:quantum}
for the quantum case. In sec. \ref{sec:Algebra} we discuss the relationship
between both linear and second-order invariants, in classical and
quantum case. This analysis agrees with Manko's and Dodonov's idea
\cite{Manko1} of the fundamental properties of the first-order integrals
of motion. In sec. \ref{sec:Final-remarks} we highlight our main
results.

\section{The classical driven TDHO\label{sec:classical}}

For the start point, let us take the second-order ODE that describes
a one-dimensional classical time-dependent harmonic oscillator with
a driven force, 
\begin{equation}
\frac{d^{2}q}{dt^{2}}+\omega^{2}\left(t\right)q=F\left(t\right),\label{eq:2.01}
\end{equation}
in which the physical parameters as mass and strength of the driving
term are normalized and absorbed by the variables. The frequency and
the force term are considered to be analytic everywhere in $t$.

We may decompose \eqref{eq:2.01} into the following first-order ODEs\begin{subequations}\label{eq:2.02}
\begin{equation}
p=\frac{dq}{dt},\label{eq:2.02a}
\end{equation}
and 
\begin{equation}
\frac{dp}{dt}=F\left(t\right)-\omega^{2}\left(t\right)q.\label{eq:2.02b}
\end{equation}
\end{subequations}These may be recognized as the canonical equations
of the driven TDHO. The most general solution of \eqref{eq:2.02}
requires two initial data, which may be the values of $q$ and $p$
at an instant of time $t=t_{0}$. The integrability of these equations
are then determined by the knowledge of two linearly independent dynamical
invariants of the system. The canonical equations themselves has all
the information necessary to find these quantities.

\subsection{Linear invariants\label{sub:Linear-invariants}}

Let us proceed with the same procedure done in \cite{MBJ}, defining
two arbitrary complex functions $\alpha\left(t\right)$ and $\beta\left(t\right)$.
Multiplying \eqref{eq:2.02a} by $\alpha$ and \eqref{eq:2.02b} by
$\beta$, then building the linear combination, yields 
\[
\beta\frac{dp}{dt}+\alpha\frac{dq}{dt}=\alpha p+\beta\left(F-\omega^{2}q\right).
\]
Isolating the total time derivative results in the expression 
\begin{equation}
\frac{d}{dt}\left(\beta p+\alpha q\right)=\left(\alpha+\frac{d\beta}{dt}\right)p+\left(\frac{d\alpha}{dt}-\omega^{2}\beta\right)q+\beta F.\label{eq:2.03}
\end{equation}

The force term in \eqref{eq:2.03} becomes a problem, since we seek
for conditions over $\alpha$ and $\beta$ so the l.h.s. can be made
zero. We may fix this defining a function $\mathcal{F}\left(\beta,t\right)$
such that 
\begin{equation}
\mathcal{F}\left(\beta,t\right)\equiv\int_{t_{0}}^{t}\beta\left(\tau\right)F\left(\tau\right)d\tau,\,\,\,\,\,\,\,\,\,\,\,\,\,\,\,\,\,\,\,\,\beta\left(t_{0}\right)=0.\label{eq:2.04}
\end{equation}
In this case we have the identity 
\begin{equation}
\beta F=\frac{d\mathcal{F}}{dt}.\label{eq:2.05}
\end{equation}
With \eqref{eq:2.05}, we may express \eqref{eq:2.03} in the form
\begin{equation}
\frac{d}{dt}\left(\beta p+\alpha q-\mathcal{F}\right)=\left(\alpha+\frac{d\beta}{dt}\right)p+\left(\frac{d\alpha}{dt}-\omega^{2}\beta\right)q.\label{eq:2.06}
\end{equation}

If the parameters $\alpha$ and $\beta$ satisfy the ODEs\begin{subequations}\label{eq:2.07}
\begin{gather}
\alpha+\frac{d\beta}{dt}=0,\label{eq:2.07a}\\
\frac{d\alpha}{dt}-\omega^{2}\beta=0,\label{eq:2.07b}
\end{gather}
\end{subequations}the polynomial 
\begin{equation}
I_{L}=\beta p+\alpha q-\mathcal{F}\left(\beta,t\right)\label{eq:2.08}
\end{equation}
is a linear dynamical invariant of the system \eqref{eq:2.02}. We
see by \eqref{eq:2.07} that $\alpha$ and $\beta$ are not independent
functions, so we may write \eqref{eq:2.08} depending only of $\beta$:\begin{subequations}\label{eq:2.09}
\begin{equation}
I_{L}=\beta p-\frac{d\beta}{dt}q-\mathcal{F}\left(\beta,t\right),\label{eq:2.09a}
\end{equation}
where $\beta$ obeys

\begin{equation}
\left(\frac{d^{2}}{dt^{2}}+\omega^{2}\right)\beta=0.\label{eq:2.09b}
\end{equation}
\end{subequations}

The most general solution of \eqref{eq:2.09b} is a complex function
$\beta$, so we actually have two linearly independent invariants\begin{subequations}\label{eq:2.10}
\begin{gather}
I_{L}=\beta p-\frac{d\beta}{dt}q-\mathcal{F}\left(\beta,t\right),\label{eq:2.10a}\\
I_{L}^{*}=\beta^{*}p-\frac{d\beta^{*}}{dt}q-\mathcal{F}\left(\beta^{*},t\right),\label{eq:2.10b}
\end{gather}
\end{subequations}where $\beta$ and $\beta^{*}$ obey the set of
ODEs 
\begin{equation}
\left(\frac{d^{2}}{dt^{2}}+\omega^{2}\right)\left(\begin{array}{c}
\beta\\
\beta^{*}
\end{array}\right)=0.\label{eq:2.11}
\end{equation}
Eqs. \eqref{eq:2.11} are the same expected for the parameter $\beta$
in the case of a simple time-dependent harmonic oscillator \cite{MBJ}.

\subsection{Quadratic invariants\label{sub:quadratic-invariants}}

We may also build quadratic invariants from the equations of motion
\eqref{eq:2.02}. Without the driving force, it would be sufficient
to build linear combinations of products of the equations of motion.
However this is not the case when the driving force is in place. Let
us observe the following products between \eqref{eq:2.02a} and \eqref{eq:2.02b}:\begin{subequations}\label{eq:2.12}
\begin{gather}
\frac{d}{dt}\left(\frac{q^{2}}{2}\right)=qp,\label{eq:2.12a}\\
\frac{d}{dt}\left(qp\right)=p^{2}-\omega^{2}q^{2}+qF,\label{eq:2.12b}\\
\frac{d}{dt}\left(\frac{p^{2}}{2}\right)=\left(F-\omega^{2}q\right)\frac{dq}{dt}.\label{eq:2.12c}
\end{gather}
\end{subequations}The r.h.s. of these equations fail to be purely
quadratic forms in the variables $\left(q,p\right)$, because of the
presence of the driving force. This situation is corrected with the
use of the equations of motion themselves: 
\[
\frac{dq}{dt}=p,\,\,\,\,\,\,\,\,\frac{dp}{dt}=\left(F-\omega^{2}q\right).
\]

Now we take a set of time-dependent functions $c_{i}=\left(c_{1},c_{2},c_{3},c_{4},c_{5}\right)$,
and build the linear combination 
\begin{align*}
c_{1}\frac{d}{dt}\left(\frac{q^{2}}{2}\right) & +c_{2}\frac{d}{dt}\left(qp\right)+c_{3}\frac{d}{dt}\left(\frac{p^{2}}{2}\right)+c_{4}\frac{dq}{dt}+c_{5}\frac{dp}{dt}=\\
 & =c_{1}qp+c_{2}\left(p^{2}-\omega^{2}q^{2}+qF\right)+c_{3}\left(F-\omega^{2}q^{2}\right)+c_{4}p+c_{5}\left(F-\omega^{2}q\right).
\end{align*}
Collecting the total time derivatives, it yields the expression 
\begin{align}
\frac{d}{dt}\left(c_{1}\frac{q^{2}}{2}\right.+ & \left.c_{2}qp+c_{3}\frac{p^{2}}{2}+c_{4}q+c_{5}p\right)=\nonumber \\
= & \left(c_{2}+\frac{1}{2}\frac{dc_{3}}{dt}\right)p^{2}+\left(\frac{1}{2}\frac{dc_{1}}{dt}-c_{2}\omega^{2}\right)q^{2}+\left(c_{1}+\frac{dc_{2}}{dt}-c_{3}\omega^{2}\right)qp\nonumber \\
 & +\left(c_{2}F+\frac{dc_{4}}{dt}-c_{5}\omega^{2}\right)q+\left(c_{3}F+c_{4}+\frac{dc_{5}}{dt}\right)p+c_{5}F.\label{eq:2.13}
\end{align}

Now we use the identity 
\begin{equation}
c_{5}F=\frac{d\mathcal{F}\left(c_{5},t\right)}{dt}=\frac{d}{dt}\left[\int_{t_{0}}^{t}c_{5}\left(\tau\right)F\left(\tau\right)d\tau\right],\,\,\,\,\,\,\,\,\,\,\, c_{5}\left(t_{0}\right)=0,\label{eq:2.14}
\end{equation}
and in this case, 
\begin{align}
\frac{d}{dt}\left[c_{1}\frac{q^{2}}{2}\right.+ & \left.c_{2}qp+c_{3}\frac{p^{2}}{2}+c_{4}q+c_{5}p-\mathcal{F}\left(c_{5},t\right)\right]=\nonumber \\
= & \left(c_{2}+\frac{1}{2}\frac{dc_{3}}{dt}\right)p^{2}+\left(\frac{1}{2}\frac{dc_{1}}{dt}-c_{2}\omega^{2}\right)q^{2}+\left(c_{1}+\frac{dc_{2}}{dt}-c_{3}\omega^{2}\right)qp\nonumber \\
 & +\left(c_{2}F+\frac{dc_{4}}{dt}-c_{5}\omega^{2}\right)q+\left(c_{3}F+c_{4}+\frac{dc_{5}}{dt}\right)p.\label{eq:2.15}
\end{align}

Therefore the second-order polynomial 
\begin{equation}
I=\frac{c_{1}}{2}q^{2}+c_{2}qp+\frac{c_{3}}{2}p^{2}+c_{4}q+c_{5}p-\mathcal{F}\left(c_{5},t\right)\label{eq:2.16}
\end{equation}
is a dynamical invariant if the equations\begin{subequations}\label{eq:2.17}
\begin{gather}
c_{2}+\frac{1}{2}\frac{dc_{3}}{dt}=0,\\
\frac{1}{2}\frac{dc_{1}}{dt}-c_{2}\omega^{2}=0,\\
c_{1}+\frac{dc_{2}}{dt}-c_{3}\omega^{2}=0,\\
c_{2}F+\frac{dc_{4}}{dt}-c_{5}\omega^{2}=0,\\
c_{3}F+c_{4}+\frac{dc_{5}}{dt}=0
\end{gather}
\end{subequations}are satisfied.

We notice that \eqref{eq:2.16} can be rewritten to depend only of
the functions $c_{3}$ and $c_{5}$. Let us rename them as $\gamma$
and $\sigma$ respectively. This can be done with the set \eqref{eq:2.17},
and results in 
\begin{equation}
I_{Q}=\left(\frac{1}{2}\frac{d^{2}\gamma}{dt^{2}}+\omega^{2}\gamma\right)\frac{q^{2}}{2}-\frac{1}{2}\frac{d\gamma}{dt}qp+\gamma\frac{p^{2}}{2}-\left(\frac{d\sigma}{dt}+\gamma F\right)q+\sigma p-\mathcal{F}\left(\sigma,t\right).\label{eq:2.18}
\end{equation}
The ODEs for $\gamma$ and $\sigma$ follow:\begin{subequations}\label{eq:2.19}
\begin{gather}
\frac{1}{2}\frac{d^{3}\gamma}{dt^{3}}+2\omega^{2}\frac{d\gamma}{dt}+\frac{d\omega^{2}}{dt}\gamma=0,\label{eq:2.19a}\\
\frac{d^{2}\sigma}{dt^{2}}+\omega^{2}\sigma=-\gamma\frac{dF}{dt}-\frac{3}{2}\frac{d\gamma}{dt}F.\label{eq:2.19b}
\end{gather}
\end{subequations}More than that, \eqref{eq:2.19a} can be integrated
to give 
\begin{equation}
\left(\frac{1}{2}\frac{d^{2}}{dt^{2}}+\omega^{2}\right)\gamma=\frac{W^{2}}{\gamma}+\frac{1}{4\gamma}\left(\frac{d\gamma}{dt}\right)^{2},\label{eq:2.20}
\end{equation}
in which $W^{2}$ is the integration constant.

Supposing a given solution of \eqref{eq:2.20}, a function $\sigma$
can be found as a solution of \eqref{eq:2.19b}. In this case, because
of the force term, two independent parameters are required to the
construction of the second-order invariant \eqref{eq:2.18}. Using
\eqref{eq:2.20} in \eqref{eq:2.18} we have 
\begin{equation}
I_{Q}=\frac{1}{2\gamma}\left[W^{2}q^{2}+\left(\frac{1}{2}\frac{d\gamma}{dt}q-\gamma p\right)^{2}\right]-\left(\frac{d\sigma}{dt}+\gamma F\right)q+\sigma p-\mathcal{F}\left(\sigma,t\right).\label{eq:2.21}
\end{equation}
The invariant \eqref{eq:2.21} is found by K. Takayama in \cite{Takayama}
using the dynamical algebra of the related hamiltonian function, and
it is also used to model charged particles in betatron accelerators.

Let us suppose that $\gamma$ is a positive real quantity, i.e. $\gamma\equiv\rho^{2}$.
In this case, \eqref{eq:2.20} becomes 
\begin{equation}
\left(\frac{d^{2}}{dt^{2}}+\omega^{2}\right)\rho=\frac{W^{2}}{\rho^{3}},\label{eq:2.22}
\end{equation}
which is an Ermakov equation for the parameter $\rho$, and 
\begin{equation}
I_{Q}=\frac{1}{2}\left[\frac{q^{2}}{\rho^{2}}+\left(\frac{d\rho}{dt}q-\rho p\right)^{2}\right]-\left(\frac{d\sigma}{dt}+\rho^{2}F\right)q+\sigma p-\mathcal{F}\left(\sigma,t\right)\label{eq:2.23}
\end{equation}
can be seen as an Ermakov-like invariant of the driven TDHO.

\section{The quantum driven TDHO\label{sec:quantum}}

Let us now deal with the quantum analogue of the problem dealt in
sec. \ref{sec:classical}. The quantum oscillator is described by
the set of variables $\left(\hat{q},\hat{p}\right)$, which are now
self-adjoint operators acting on elements of a Hilbert space. The
equations of motion in this case become Heisenberg equations for these
operators, and their functional form is identical compared to the
classical system, i.e., 
\begin{equation}
\hat{p}=\frac{d\hat{q}}{dt},\,\,\,\,\,\,\,\,\,\,\,\,\,\,\frac{d\hat{p}}{dt}=F\left(t\right)-\omega^{2}\left(t\right)\hat{q}.\label{eq:3.01}
\end{equation}
The operators $\hat{q}$ and $\hat{p}$ do not commute, i.e. $\left[\hat{q},\hat{p}\right]\equiv\hat{q}\hat{p}-\hat{p}\hat{q}\neq0$,
therefore the ordering problem between these operators has to be taken
into account. The force term is understood to be the unity operator
multiplied by an analytic time dependent function $F\left(t\right)$.

\subsection{Quantum invariants\label{sub:Quantum-invariants}}

Since the ordering problem does not affect linear combinations of
the eqs. \eqref{eq:3.01}, the same construction of sec. \ref{sub:Linear-invariants}
can be made. The result is simply that the operators\begin{subequations}\label{eq:3.02}
\begin{gather}
\hat{I}_{L}=\beta\hat{p}-\frac{d\beta}{dt}\hat{q}-\mathcal{F}\left(\beta,t\right),\label{eq:3.02a}\\
\hat{I}_{L}^{\dagger}=\beta^{*}\hat{p}-\frac{d\beta^{*}}{dt}\hat{q}-\mathcal{F}\left(\beta^{*},t\right),\label{eq:3.02b}
\end{gather}
\end{subequations}in which 
\begin{equation}
\mathcal{F}\left(\beta,t\right)\equiv\int_{t_{0}}^{t}\beta\left(\tau\right)F\left(\tau\right)d\tau,\,\,\,\,\,\,\,\,\beta\left(t_{0}\right)=0,\label{eq:3.03}
\end{equation}
are linear invariants of \eqref{eq:3.01} if $\beta$ and $\beta^{*}$
obey the equations 
\begin{equation}
\left(\frac{d^{2}}{dt^{2}}+\omega^{2}\right)\left(\begin{array}{c}
\beta\\
\beta^{*}
\end{array}\right)=0,\label{eq:3.04}
\end{equation}
which are the same eqs. \eqref{eq:2.11}.

The construction of the quadratic dynamical invariants, on the other
hand, needs special attention to the ordering problem. Taking \eqref{eq:3.01},
we may build the products\begin{subequations}\label{eq:3.05} 
\begin{gather}
\frac{d\hat{q}^{2}}{dt}=\left\{ \hat{q},\hat{p}\right\} ,\\
\frac{d}{dt}\left\{ \hat{q},\hat{p}\right\} =2\hat{p}^{2}+2F\hat{q}-2\omega^{2}\hat{q}^{2},\\
\frac{d\hat{p}^{2}}{dt}=2F\hat{p}-\omega^{2}\left\{ \hat{q},\hat{p}\right\} ,
\end{gather}
\end{subequations}where $\left\{ \hat{q},\hat{p}\right\} \equiv\hat{q}\hat{p}+\hat{p}\hat{q}$
is the anti-commutator. Once again the presence of the force term
requires the use of the Heisenberg equations themselves: 
\begin{gather*}
\frac{d\hat{q}}{dt}=\hat{p},\;\;\;\;\;\frac{d\hat{p}}{dt}=F-\omega^{2}\hat{q}.
\end{gather*}

We define a set of time-dependent functions $c_{i}=\left(c_{1},c_{2},c_{3},c_{4},c_{5}\right)$,
and build the linear combination between these equations. Taking the
total time-derivative gives the expression 
\begin{align}
\frac{d}{dt}\left[c_{1}\frac{\hat{q}^{2}}{2}\right.+ & \left.\frac{1}{2}c_{2}\left\{ \hat{q},\hat{p}\right\} +c_{3}\frac{\hat{p}^{2}}{2}+c_{4}\hat{q}+c_{5}\hat{p}-\mathcal{F}\left(c_{5},t\right)\right]=\nonumber \\
= & \left(c_{2}+\frac{1}{2}\frac{dc_{3}}{dt}\right)\hat{p}^{2}+\left(\frac{1}{2}\frac{dc_{1}}{dt}-c_{2}\omega^{2}\right)\hat{q}^{2}+\frac{1}{2}\left(c_{1}+\frac{dc_{2}}{dt}-c_{3}\omega^{2}\right)\left\{ \hat{q},\hat{p}\right\} \nonumber \\
 & +\left(c_{2}F+\frac{dc_{4}}{dt}-c_{5}\omega^{2}\right)\hat{q}+\left(c_{3}F+c_{4}+\frac{dc_{5}}{dt}\right)\hat{p}.\label{eq:3.06}
\end{align}
Using \eqref{eq:3.03}, we reach the quantum second-order polynomial
\begin{equation}
\hat{I}_{Q}=c_{1}\frac{\hat{q}^{2}}{2}+\frac{1}{2}c_{2}\left\{ \hat{q},\hat{p}\right\} +c_{3}\frac{\hat{p}^{2}}{2}+c_{4}\hat{q}+c_{5}\hat{p}-\mathcal{F}\left(c_{5},t\right),\label{eq:3.07}
\end{equation}
which is a dynamical invariant if the set\begin{subequations}\label{eq:3.08}
\begin{gather}
c_{2}+\frac{1}{2}\frac{dc_{3}}{dt}=0,\\
\frac{1}{2}\frac{dc_{1}}{dt}-c_{2}\omega^{2}=0,\\
c_{1}+\frac{dc_{2}}{dt}-c_{3}\omega^{2}=0,\\
c_{2}F+\frac{dc_{4}}{dt}-c_{5}\omega^{2}=0,\\
c_{3}F+c_{4}+\frac{dc_{5}}{dt}=0
\end{gather}
\end{subequations}is satisfied. These are the same equations \eqref{eq:2.18}.

Again, using \eqref{eq:3.08} it is possible to express \eqref{eq:3.07}
in terms of $c_{3}\equiv\gamma$ and $c_{5}\equiv\sigma$: 
\begin{equation}
\hat{I}_{Q}=\left(\frac{1}{2}\frac{d^{2}\gamma}{dt^{2}}+\omega^{2}\gamma\right)\frac{\hat{q}^{2}}{2}-\frac{1}{4}\frac{d\gamma}{dt}\left\{ \hat{q},\hat{p}\right\} +\gamma\frac{\hat{p}^{2}}{2}-\left(\frac{d\sigma}{dt}+\gamma F\right)\hat{q}+\sigma\hat{p}-\mathcal{F}\left(\sigma,t\right).\label{eq:3.09}
\end{equation}
The equations for $\gamma$ and $\sigma$ are also the same of the
classical system:\begin{subequations}\label{eq:3.10} 
\begin{gather}
\frac{1}{2}\frac{d^{3}\gamma}{dt^{3}}+2\omega^{2}\frac{d\gamma}{dt}+\frac{d\omega^{2}}{dt}\gamma=0,\\
\frac{d^{2}\sigma}{dt^{2}}+\omega^{2}\sigma=-\gamma\frac{dF}{dt}-\frac{3}{2}\frac{d\gamma}{dt}F.
\end{gather}
\end{subequations}The $\gamma$ equation can be again integrated
to give the second-order ODE \eqref{eq:2.20}. Eqs. \eqref{eq:2.21},
\eqref{eq:2.22}, and \eqref{eq:2.23} also follow straightforwardly.

\section{Algebra of the dynamical invariants\label{sec:Algebra}}

\subsection{Classical case}

In sec. \ref{sec:classical} we see that the linear invariants for
the driven TDHO are given by\begin{subequations}\label{eq:4.01}
\begin{gather}
I_{L}=\beta p-\frac{d\beta}{dt}q-\mathcal{F}\left(\beta,t\right),\\
I_{L}^{*}=\beta^{*}p-\frac{d\beta^{*}}{dt}q-\mathcal{F}\left(\beta^{*},t\right),
\end{gather}
\end{subequations}provided that $\beta$ and $\beta^{*}$ obey 
\begin{equation}
\left(\frac{d^{2}}{dt^{2}}+\omega^{2}\right)\left(\begin{array}{c}
\beta\\
\beta^{*}
\end{array}\right)=0,\label{eq:4.02}
\end{equation}
and 
\begin{equation}
\mathcal{F}\left(f,t\right)=\int_{t_{0}}^{t}f\left(\tau\right)F\left(\tau\right)d\tau,\,\,\,\,\,\,\,\, f\left(t_{0}\right)=0,\label{eq:4.03}
\end{equation}
for a generic function $f\left(t\right)$.

On the other hand, the quadratic form 
\begin{equation}
I_{Q}=\left(\frac{1}{2}\frac{d^{2}\gamma}{dt^{2}}+\omega^{2}\gamma\right)\frac{q^{2}}{2}-\frac{1}{2}\frac{d\gamma}{dt}qp+\gamma\frac{p^{2}}{2}-\left(\frac{d\sigma}{dt}+\gamma F\right)q+\sigma p-\mathcal{F}\left(\sigma,t\right)\label{eq:4.04}
\end{equation}
is a second-order invariant of the driven TDHO if $\gamma$ and $\sigma$
obey the set\begin{subequations}\label{eq:4.05} 
\begin{gather}
\frac{1}{2}\frac{d^{3}\gamma}{dt^{3}}+2\omega^{2}\frac{d\gamma}{dt}+\frac{d\omega^{2}}{dt}\gamma=0,\label{eq:4.05a}\\
\frac{d^{2}\sigma}{dt^{2}}+\omega^{2}\sigma=-\gamma\frac{dF}{dt}-\frac{3}{2}\frac{d\gamma}{dt}F.\label{eq:4.05b}
\end{gather}
\end{subequations}We now ask the question if the quadratic invariant
can be related with the linear ones.

Let us consider the product 
\begin{align}
I_{L}^{*}I_{L}= & \frac{d\beta^{*}}{dt}\frac{d\beta}{dt}q^{2}-\frac{d}{dt}\left(\beta^{*}\beta\right)qp+\left(\beta^{*}\beta\right)p^{2}+\left[\frac{d\beta^{*}}{dt}\mathcal{F}\left(\beta,t\right)+\frac{d\beta}{dt}\mathcal{F}\left(\beta^{*},t\right)\right]q\nonumber \\
 & -\left[\beta^{*}\mathcal{F}\left(\beta,t\right)+\beta\mathcal{F}\left(\beta^{*},t\right)\right]p+\left|\mathcal{F}\left(\beta,t\right)\right|^{2}.\label{eq:4.06}
\end{align}
Any product of invariants is also an invariant. In the case of \eqref{eq:4.06},
it becomes a second-order real invariant in the variables $p$ and
$q$. Observing \eqref{eq:4.04}, both quadratic forms are equivalent
if\begin{subequations}\label{eq:4.07} 
\begin{gather}
\gamma=2\beta^{*}\beta,\label{eq:4.07a}\\
\frac{1}{2}\left(\frac{1}{2}\frac{d^{2}\gamma}{dt^{2}}+\omega^{2}\gamma\right)=\frac{d\beta^{*}}{dt}\frac{d\beta}{dt},\label{eq:4.07b}\\
\frac{d\sigma}{dt}+\gamma F=-\frac{d\beta^{*}}{dt}\mathcal{F}\left(\beta,t\right)-\frac{d\beta}{dt}\mathcal{F}\left(\beta^{*},t\right),\label{eq:4.07c}\\
\sigma=-\beta^{*}\mathcal{F}\left(\beta,t\right)-\beta\mathcal{F}\left(\beta^{*},t\right),\label{eq:4.07d}\\
\mathcal{F}\left(\sigma,t\right)=-\left|\mathcal{F}\left(\beta,t\right)\right|^{2}.\label{eq:4.07e}
\end{gather}
\end{subequations}are satisfied.

Let us observe the term on the l.h.s. of \eqref{eq:4.07b}. Using
\eqref{eq:4.07a} we have 
\[
\frac{1}{2}\left(\frac{1}{2}\frac{d^{2}}{dt^{2}}+\omega^{2}\right)\gamma=\frac{1}{2}\left(\frac{d^{2}\beta^{*}}{dt^{2}}+\omega^{2}\beta^{*}\right)\beta+\frac{1}{2}\left(\frac{d^{2}\beta}{dt^{2}}+\omega^{2}\beta\right)\beta^{*}+\frac{d\beta^{*}}{dt}\frac{d\beta}{dt}.
\]
Since $\beta$ and $\beta^{*}$ obey \eqref{eq:4.02}, \eqref{eq:4.07b}
is identically satisfied.

Now we take \eqref{eq:4.07d} and substitute $\sigma$ in the l.h.s.
of eq. \eqref{eq:4.07c}. The result is 
\[
\frac{d\sigma}{dt}+\gamma F=-\frac{d\beta^{*}}{dt}\mathcal{F}\left(\beta,t\right)-\frac{d\beta}{dt}\mathcal{F}\left(\beta^{*},t\right),
\]
which is exactly \eqref{eq:4.07c}. On the other hand, \eqref{eq:4.07d}
implies 
\[
\mathcal{F}\left(\sigma,t\right)=-\left|\mathcal{F}\left(\beta,t\right)\right|^{2}.
\]
Then, \eqref{eq:4.07e} is also identically satisfied.

We still have to show that \eqref{eq:4.07a} and \eqref{eq:4.07d}
are solutions of \eqref{eq:4.05}. In the case of \ref{eq:4.05a},
it is more convenient to work with the equivalent equation 
\begin{equation}
\frac{d}{dt}\left[\frac{1}{2}\gamma\frac{d^{2}\gamma}{dt^{2}}+\omega^{2}\gamma^{2}-\frac{1}{4}\left(\frac{d\gamma}{dt}\right)^{2}\right]=0.\label{eq:4.08}
\end{equation}
Inside the brackets, using \eqref{eq:4.02} and \eqref{eq:4.07a},
we have 
\[
\frac{1}{2}\gamma\frac{d^{2}\gamma}{dt^{2}}+\omega^{2}\gamma^{2}-\frac{1}{4}\left(\frac{d\gamma}{dt}\right)^{2}=-\left(\frac{d\beta^{*}}{dt}\beta-\beta^{*}\frac{d\beta}{dt}\right)^{2}.
\]
The quantity 
\begin{equation}
W\left(\beta^{*},\beta\right)=\frac{d\beta^{*}}{dt}\beta-\beta^{*}\frac{d\beta}{dt}\label{eq:4.09}
\end{equation}
is the Wronskian of the functions $\beta^{*}$ and $\beta$. From
\eqref{eq:4.02}, it is straightforward to show that the Wronskian
\eqref{eq:4.09} is a constant of motion. Therefore, if \eqref{eq:4.02}
is true, \eqref{eq:4.08} is identically satisfied. In fact the integration
constant in \eqref{eq:2.20} is precisely the square of the Wronskian
\eqref{eq:4.09}.

Let us see what happens with \eqref{eq:4.05b}. The l.h.s. yields
\begin{align*}
\left(\frac{d^{2}}{dt^{2}}+\omega^{2}\right)\sigma= & -\mathcal{F}\left(\beta,t\right)\left(\frac{d^{2}}{dt^{2}}+\omega^{2}\right)\beta^{*}-\mathcal{F}\left(\beta^{*},t\right)\left(\frac{d^{2}}{dt^{2}}+\omega^{2}\right)\beta\\
 & -\frac{d}{dt}\left(\beta^{*}\beta\right)F-2\frac{d}{dt}\left(\beta^{*}\beta F\right).
\end{align*}
Again using \eqref{eq:4.02}, 
\[
\left(\frac{d^{2}}{dt^{2}}+\omega^{2}\right)\sigma=-\gamma\frac{dF}{dt}-\frac{3}{2}\frac{d\gamma}{dt}F,
\]
where \eqref{eq:4.07a} is used. This reproduces eq. \eqref{eq:4.05b},
as required.

\subsection{Quantum case}

Due to the Heisenberg algebra of the quantum driven TDHO, it is necessary
to consider symmetric and antisymmetric products of the linear invariants
\eqref{eq:3.02}. The antisymmetric product is given by 
\[
\hat{I}_{A}=\frac{1}{2}\left[\hat{I}_{L}^{\dagger},\hat{I}_{L}\right]=-\frac{1}{2}W\left(\beta^{*},\beta\right)\left[\hat{q},\hat{p}\right],
\]
which is not a dynamical invariant, since it is a pure constant term.
On the other hand, the symmetric product 
\begin{align}
\hat{I}_{S}= & \frac{1}{2}\left\{ \hat{I}_{L}^{\dagger},\hat{I}_{L}\right\} =\beta^{*}\beta\hat{p}^{2}-\frac{1}{2}\frac{d}{dt}\left(\beta^{*}\beta\right)\left\{ \hat{q},\hat{p}\right\} +\frac{d\beta^{*}}{dt}\frac{d\beta}{dt}\hat{q}^{2}\nonumber \\
 & -\left[\beta^{*}\mathcal{F}\left(\beta,t\right)+\mathcal{F}\left(\beta^{*},t\right)\beta\right]\hat{p}+\left[\frac{d\beta^{*}}{dt}\mathcal{F}\left(\beta,t\right)+\mathcal{F}\left(\beta^{*},t\right)\frac{d\beta}{dt}\right]\hat{q}+\left|\mathcal{F}\left(\beta,t\right)\right|^{2}\label{eq:4.10}
\end{align}
is a true quadratic invariant.

Let us seek for the conditions that allow \eqref{eq:4.10} to be equal
to \eqref{eq:3.09}. We find that the following set of equations:\begin{subequations}\label{eq:4.11}
\begin{gather}
\gamma=2\beta^{*}\beta,\\
\sigma=-\beta^{*}\mathcal{F}\left(\beta,t\right)-\mathcal{F}\left(\beta^{*},t\right)\beta,
\end{gather}
\end{subequations}together with\begin{subequations}\label{eq:4.12}
\begin{gather}
\frac{1}{2}\frac{d^{2}\gamma}{dt^{2}}+\gamma\omega^{2}=2\frac{d\beta^{*}}{dt}\frac{d\beta}{dt},\\
\frac{d\sigma}{dt}+\gamma F=-\frac{d\beta^{*}}{dt}\mathcal{F}\left(\beta,t\right)-\mathcal{F}\left(\beta^{*},t\right)\frac{d\beta}{dt},\\
\mathcal{F}\left(c_{5},t\right)=-\left|\mathcal{F}\left(\beta,t\right)\right|^{2}
\end{gather}
\end{subequations}are required. As for the classical case, eqs. \eqref{eq:4.11}
implies \eqref{eq:4.12}. Also, \eqref{eq:4.11} are solutions of
\eqref{eq:3.10} if $\beta$ and $\beta^{*}$ are solutions of \eqref{eq:3.04}.

\section{Final remarks\label{sec:Final-remarks}}

In this paper we analysed the problem of construction of linear and
quadratic dynamical invariants for the classical and quantum driven
time-dependent harmonic oscillator. The procedure, applied in \cite{MBJ}
for the case of the unforced harmonic oscillator, requires nothing
more than the first-order equations of motion (or the Heisenberg equations
in the quantum case). As for the unforced case, the linear dynamical
invariants \eqref{eq:2.10} (\eqref{eq:3.02} in the quantum case)
of the driven TDHO are built from linear combinations of the equations
of motion \eqref{eq:2.02} (\eqref{eq:3.01}), provided the complex
parameter $\beta$ and its conjugated $\beta^{*}$ obey the ODEs \eqref{eq:2.11}.
We observed that the presence of the driven force does not modify
the equations for $\beta$, although the functional form of the linear
invariants is modified by the introduction of the function $\mathcal{F}\left(\beta,t\right)$.

In the case of quadratic invariants \eqref{eq:2.18} (\eqref{eq:3.09}),
they can be constructed as linear combinations of quadratic products
of the first-order equations of motion, but because of the existence
of the driven force, they also need the contribution of simple linear
combinations of the first-order equations themselves. Two independent
parameters $\gamma$ and $\sigma$ are required to obey eqs. \eqref{eq:2.19}.
While $\sigma$ does not exist in the unforced problem, the equation
for $\gamma$ is exactly the one find in \cite{MBJ}. The existence
of the driven force yields, in this case, a second-order invariant
which is not a pure quadratic form, but depends linearly of the variables
$q$ and $p$ (or the respective Hilbert space operators in the quantum
case).

The most interesting result in this method is the fact that the linear
invariants can be used as building blocks for the construction of
second-order invariants. This is illustrated in \cite{MBJ} for the
TDHO, and now in sec. \ref{sec:Algebra} for the driven TDHO. The
second-order invariant \eqref{eq:4.04} (\eqref{eq:3.09}) is achieved
from products of the linear invariants. In the classical case, the
necessary and sufficient conditions for the product $I_{L}^{*}I_{L}$
to be equal to the quadratic invariant \eqref{eq:4.04} is that $I_{L}$
and $I_{L}^{*}$ must be linear dynamical invariants, and the parameters
$\gamma$ and $\sigma$ must be related to $\beta$ and $\beta^{*}$
through eqs. \eqref{eq:4.07a} and \eqref{eq:4.07d}. In the quantum
case, one must consider the symmetric product \eqref{eq:4.10} between
$\hat{I}_{L}$ and $\hat{I}_{L}^{\dagger}$, and the result is that
the same theorem of the classical case is proved. This result agrees
with Man'ko's and Dodonov's statement \cite{Manko1}, that linear
invariants are the fundamental quantities related to the quantization
of dynamical systems.

The proposed method allows the study of the classical and quantum
systems dynamics without using either hamiltonian or lagrangian formulations,
since we need only the equations of motion to study the symmetries
and construct the \emph{algebrae} associated to the dynamics of the
system. Following this idea, it is possible the study of systems in
which the definition of such dynamical functions is not well defined;
as far as we know, the study of the symmetries of any system in which
the dynamical equations are linear. As an example, we mention systems
with dissipative interactions up to second-order in the velocities.
Those physical systems, highly studied in the analysis of more realistic
systems, have a problematic definition of the hamiltonian or lagrangian
functions and the construction of the quantization procedures.

\subsection*{Acknowledgements}

The authors would like to thank to the Professor Dieter Schuch for
his careful reading and observations. M.C.B. thanks to FAPESP for
partial support, B.M.P. thanks CAPES and CNPq for partial support
and J.A.Ramirez thanks CAPES for full support.

\end{document}